# BioBricks.ai: A Versioned Data Registry for Life Sciences Data Assets

Author(s):

Yifan Gao[1], Zakariyya Mughal[2], Jose A. Jaramillo-Villegas[3,4], Marie Corradi[5], Alexandre Borrel[6], Ben Lieberman[2], Suliman Sharif[2], John Shaffer[2], Karamarie Fecho[7,8], Ajay Chatrath[9], Alexandra Maertens[1], Marc A.T. Teunis[5], Nicole Kleinstreuer[10], Thomas Hartung[1, 11], Thomas Luechtefeld[1,2*]

[1]Center For Alternative to Animal Testing, Johns Hopkins University, Baltimore, MD, USA

[2]Insilica, Bethesda, MD, USA

[3]Laboratory for Research in Complex Systems, Menlo Park, California, USA

[4]Facultad de Ingenierías, Universidad Tecnológica de Pereira, Pereira, Colombia

[5]Innovative Testing in Life Sciences & Chemistry, University of Applied Sciences Utrecht, Utrecht, The Netherlands

[6]Inotiv, Research Triangle Park, North Carolina, USA

[7]Renaissance Computing Institute, University of North Carolina at Chapel Hill, Chapel Hill, North Carolina, USA

[8]Copperline Professional Solutions, LLC, Pittsboro, NC, USA

[9]Department of Neurological Surgery, Washington University in Saint Louis, Saint Louis, Missouri

[10]NTP Interagency Center for the Evaluation of Alternative Methods, Research Triangle Park, NC, USA

[11]University of Konstanz, Germany

**\* Correspondence:**
Thomas Luechtefeld
tom@insilica.co

**Abstract**

Researchers in biomedical research, public health and the life sciences often spend weeks or months discovering, accessing, curating, and integrating data from disparate sources, significantly delaying the onset of actual analysis and innovation. Instead of countless developers creating redundant and inconsistent data pipelines, BioBricks.ai offers a centralized data repository and a suite of developer-friendly tools to simplify access to scientific data. Currently, BioBricks.ai delivers over ninety biological and chemical datasets. It provides a package manager-like system for installing and managing dependencies on data sources. Each 'brick' is a Data Version Control git repository that supports an updateable pipeline for extraction, transformation, and loading data into the BioBricks.ai backend at https://biobricks.ai. Use cases include accelerating data science workflows and facilitating the creation of novel data assets by integrating multiple datasets into unified, harmonized resources. In conclusion, BioBricks.ai offers an opportunity to accelerate access and use of public data through a single open platform.

1. **Introduction**

The integration of artificial intelligence (AI) into toxicology and biochemistry is revolutionizing these fields, enhancing data analysis capabilities and contributing to more efficient and accurate insights. AI methods excel at processing large, diverse datasets, which are increasingly valuable for modern toxicology and biochemistry research (1,2). In toxicology, AI-powered predictive tools like Read-Across Structure-Activity Relationships (RASAR) have achieved 87% balanced accuracy across nine Organisation for Economic Co-operation and Development (OECD) tests and 190,000 chemicals, surpassing traditional methods in predicting chemical toxicity (3). Large language models are making a growing impact on chemistry, with the capacity to predict chemical properties, evaluate synthesis pathways, and generate compounds optimized to reduce toxicity (4). These models require and benefit from large amounts of data, but many of the same datasets used for these assets are laboriously collected repeatedly from different research groups(5–7). The power of AI in fields such as toxicology and biochemistry depends heavily on the quality, quantity, and accessibility of data. Standardized data access is essential for integrating diverse data types, ensuring reproducibility, and facilitating the training and validation of AI models. Standardization supports cross-disciplinary research, regulatory compliance, and efficiency by reducing the time researchers spend on data preparation. The lack of large, high-quality training datasets is a critical barrier to the broader application of AI models in fields such as public health (8). The demand for data often surpasses the pace at which new datasets are generated and made available, highlighting the need for better data collection, management, and sharing practices (9).

There are many independent databases for public health. The European Bioinformatics Institute's identifiers.org, a registry for biomedical datasets, lists 838 such distinct data sources. This is by no means an exhaustive list, but illustrates the diverse landscape of available public health information (10,11). A survey of data scientists performed in 2022 reported that about 38% of developer effort is spent on accessing and cleaning data, rather than modeling and analyzing it (8), thus wasting valuable resources and working hours.

BioBricks solves the problem by providing a package manager for data. It provides a standardized format that works well with developer tools and allows users to have a single location to search for and install data assets.By streamlining data management and distribution, BioBricks.ai has the potential to accelerate the pace of progress in the life sciences. It reduces barriers to data access, collaboration, and distribution, allowing researchers to focus on analysis and innovation rather than data preparation and management. Herein, we provide a detailed overview of BioBricks and describe several application use cases.

## 2. Methods
### 2.1. BioBricks.ai Overview

BioBricks.ai aims to simplify the provisioning of this training and evaluation data. With a few lines of code, datasets can be loaded into a computation environment. BioBricks.ai provides a public, centralized Data Version Control (DVC) (https://dvc.org/doc/use-cases/data-registry) data registry for public health data assets (13,14). While built on DVC for data science projects, BioBricks.ai enhances this foundation with a specialized command-line tool and web portal focused on installing and managing data dependencies in a manner akin to package managers like the Comprehensive R Archive Network (CRAN), Bioconductor and PyPI (15,16).

BioBricks.ai manages data assets organized into 'brick's. Each brick is a git repository adhering to a standardized protocol outlined in the BioBricks.ai template repository (github.com/biobricks-ai/brick-template). Bricks can be created with or without dependencies on other bricks. For independent bricks, which often represent primary data sources, BioBricks.ai's policy is to replicate the original data without modifications, ensuring data integrity and fidelity, with full attribution and citation. Examples include the HUGO Gene Nomenclature Committee brick (github.com/biobricks-ai/hgnc) and the ClinVar brick (github.com/biobricks-ai/clinvar), a database of clinical variants and their relationship to human health, (17,18).

Bricks can also be built with dependencies on other bricks, like these primary sources, allowing for more complex data structures that might restructure data, combine multiple sources, or generate derived products like machine learning models. This flexible structure enables BioBricks.ai to maintain a hierarchy of data resources, from raw datasets to sophisticated, integrated products. A prime example is the ChemHarmony brick, which combines and simplifies data from over fifteen chemical-safety–related databases into a single, unified schema using unique, curated chemical identifiers. By providing standardized access to consistent versions of datasets, BioBricks.ai significantly reduces data acquisition time, facilitates collaboration among researchers, and simplifies the process of building downstream assets that depend on multiple upstream data sources.

With a straightforward installation process, the BioBricks.ai tool offers a unified interface to discover and utilize numerous data sources. Instead of navigating multiple databases, APIs, packages, or specialized data tools for each new source, researchers only need to learn one straightforward system. The accompanying web application, https://biobricks.ai, enables tracking of asset usage, potentially facilitating future features like bandwidth cost allocation and enhanced tooling around constructed data sources.

BioBricks.ai can be used to quickly install 'bricks', which are git repositories with code for building databases (or other data assets). Getting set up involves installing the command line tool, configuring the tool, and then installing bricks:

Code 1 - configuring BioBricks and installing a brick can be done in 3 steps

```bash
pipx install biobricks
biobricks configure
biobricks install <brickname> # eg `biobricks install hgnc`
```

BioBricks.ai recommends using pipx to install the command line tool in an isolated environment. Pipx is a command-line utility that enables users to install Python packages into isolated environments. By using pipx to install the BioBricks command line tool, users can run commands such as biobricks install without the need to manage dependency conflicts with other Python environments. Alternatively, users can install BioBricks.ai using pip install biobricks if preferred. The tool is designed to be lightweight, with minimal dependencies, ensuring a simple and efficient installation process. Importantly, while DVC is used in the development of bricks, it is not required for end-users of the BioBricks.ai tool, further streamlining the user experience for researchers who need only to access and utilize the data.

Researchers can find all available bricks developed by the BioBricks.ai team on GitHub at github.com/biobricks-ai and on the official website at https://biobricks.ai. To use the tool, users create an account at BioBricks.ai, which provides them with a token that enables asset downloads. Several example bricks are shown in Table 1. Bricks are categorized based on the data they contain: Chemical Informatics, Cancer Research, Genomics & Genetics, Proteomics, Pharmacology and Drug Discovery, Toxicology and Environmental Science, Medical and Clinical Sciences, Ontology and Terminology, and Systems Biology and Pathways. A knowledge graph visualization of each brick and their associated category is shown below in

Table 1 - Examples of Databases in BioBricks.ai

| Repository | Description |
| --- | --- |
| **SMRT** | Small Molecule Retention Time Dataset. |
| **dictrank** | Drug-Induced Cardiotoxicity Rank Dataset. |
| **ice** | Integrated Chemical Environment - High quality in vitro and in vivo toxicology data. |
| **biogrid** | Data from BioGRID. |
| **ctgov** | Data from ClinicalTrials.gov. |
| **mirbase** | Data from miRBase. |
| **skinsensdb** | Skin sensitization database. |
| **ctdbase** | Data from Comparative Toxicogenomics Database. |
| **tox21** | Tox21 quantitative high throughput screening (qHTS) 10K library data. |
| **targetscan** | Data from TargetScan. |
| **USPTO_ChemReaction** | Data from USPTO Chemical Reaction Database. |
| **moleculenet** | Molecular datasets for machine learning. |
| **pubchem** | PubChem data. |
| **toxvaldb** | Toxicity endpoint data. |
| **dbgap** | Genotype-phenotype interaction data. |
| **zinc** | ZINC purchasable compound database. |
| **toxcast** | EPA in vitro toxicity data. |
| **pdb** | Protein Data Bank 3D structure data. |
| **geneontology** | Gene Ontology knowledgebase. |
| **cpdat** | Consumer Product Data. |
| **cpcat** | Chemical Product Categories. |
| **chembl** | Bioactive molecule data. |

Documentation for building, installing, and configuring BioBricks.ai is available at docs.biobricks.ai. In the following sections, we provide some details on how BioBricks.ai functions, but we refer active users to the documentation for complete details.

## 2.2. BioBricks.ai Configuration & Architecture

After installing biobricks with pipx install biobricks, users can run biobricks configure to set up a BioBricks.ai library, which is a directory located at a user-defined path with a special cache subdirectory named ./cache. When users install bricks, they are stored in subdirectories within the BioBricks library.

The cache directory stores data files, each uniquely identified by a content-based MD5 hash. This hashing mechanism ensures that each distinct data file is stored only once, minimizing data duplication, optimizing storage efficiency, and enhancing data retrieval processes. When multiple bricks or different versions of the same brick require access to an identical data file, the file is retrieved from the cache, reducing redundancy and conserving disk space.

The git directories in the biobricks library are organized into subdirectories identified by an owner, assigned name, and commit hash ./{orgname}/{reponame}/{commit-hash}. For instance, a repository path might be ./biobricks-ai/chemharmony/4f060. This format separates repositories by organization and name, while concurrently storing specific versions of each repository. This hierarchical structure supports version control and reproducibility, facilitating navigation between different repository versions. BioBricks.ai enforces a default organization of github.com/biobricks-ai when using the briobricks install command. However, other URLs can be referenced by providing the full git URL and commit hash.

The integration of a content-based hash system allows multiple bricks or versions to reference the same data file without unnecessary duplication. The git structure simplifies the management of various brick versions. Collectively, these systems provide an organized framework for the management of bioinformatics data and code repositories, enhancing both storage efficiency and data accessibility.

### 2.3. Brick Installation

Bricks can be installed using the command biobricks install <brickname>, which downloads bricks from the default biobricks-ai organization. Bricks can also be installed using the full URL for a specific git repository, followed by a specific commit hash; this allows non-BioBricks.ai developers to build their own bricks.

After issuing the install command, the BioBricks.ai system 1) clones the relevant git repository to the user's library, 2) looks up all of the assets in the repository's 'brick' directory, 3) downloads those assets to the repository's brick directory from Amazon S3 to the users cache directory, and 4) builds symlinks from the brick directory to the files in the cache directory. When the files already exist in the cache directory, step 3 is skipped.

To illustrate, consider the process of installing a brick from BioBricks.ai, named HGNC, using a specific commit 4f060 (if a user does not specify a commit the most recent commit on the main branch is used). BioBricks.ai first checks if this version of the brick already exists in the library. If not, it authenticates the user, checks if the URL is a functional git repository, and (if these steps succeed) clones the repository into the directory ./biobricks-ai/HGNC/4f060. BioBricks.ai then fetches all the data from the HGNC ./brick directory and stores it in the ./cache directory (file cache paths are generated from their hash values). After all steps are completed, BioBricks.ai logs a success message, indicating that the brick is ready for use.

## 2.4. Accessing Data with BioBricks.ai

After installing the hgnc brick using biobricks install hgnc, the command line tool biobricks assets lists the objects (tables, databases, semantic graphs) in hgnc repositories' brick directory. In Python and R, you can use the BioBricks.ai library to access this data in a developer-friendly manner.

Code 2 - accessing a database in BioBricks.ai

```python
Python

import biobricks as bb
import pandas as pd
hgnc = bb.assets('hgnc')
pd.read_parquet(hgnc.hgnc_complete_set_parquet)
```

The bb.assets('hgnc') function returns a SimpleNamespace, where each attribute corresponds to a specific asset within the hgnc brick directory. This design provides a convenient way to access the files using named references.

For instance, the hgnc brick contains a parquet file named **hgnc_complete_set_parquet** the path for which is accessed with bb.assets('hgnc').hgnc_complete_set_parquet.

## 2.5. What is a Brick?

A BioBrick git repository contains user-developed code, dvc.yaml and dvc.lock files, an optional .bb directory, and a ./brick directory. The brick directory stores the developed artifacts such as parquet, sqlite, and HDT files. The dvc.yaml file describes the workflow for building the brick, while the dvc.lock file contains content-based hashes for the generated data files. The .bb directory includes a dependencies.txt file with references to data dependencies and other BioBricks.ai-specific files, like LinkML descriptions.

## 2.6. Building a Brick

Users typically start a new brick by cloning the BioBricks.ai-template repository located at github.com/biobricks-ai/brick-template. This template repo provides a project skeleton containing all the components necessary to build a brick. Bricks are built by using the dvc command line tool and running dvc repro; this command manages dependencies between code stages and runs (or reruns) for any stages required for building the brick. These stages are referenced in the dvc.yaml file present in every brick, a template for which is provided in biobricks-ai/brick-template. Eventually, metadata for every brick will be defined in the brick's linkml file (see Research Directions).

## 2.7. SMRT Example

The SMRT (Small Molecule Retention Time) brick, available at [github.com/biobricks-ai/SMRT](github.com/biobricks-ai/SMRT), serves as an excellent introductory example. Users can visit this github page and open a codespace to access a preloaded development environment for building a brick. This example illustrates how BioBricks.ai uses DVC to manage extract-transform-load (ETL) pipelines that pull data from a source, transform that data into a BioBricks.ai supported format, and load it into the BioBricks.ai backend. It will help readers grasp the practical implementation of BioBricks.ai's data management approach.

DVC manages repositories with a dvc.yaml file containing stages. Each stage has a command, dependencies, and outputs. Stages run whenever their dependencies change. The dvc.lock file maps every output and dependency to an MD5 hash, which is committed to the git history to track changes.

The SMRT brick has a simple dvc.yaml file that describes a 3-step extraction pipeline:

- **Step 1- Status**: Checks the primary source for changes.
    - **Dependencies**: None - runs on every brick update.
    - **Output**: Stores status.txt with HTML from the primary source.

- **Step 2 - Download**: Downloads raw data from the primary source to a local directory.
    - **Dependencies**: status.txt - runs when the status changes.
    - **Output**: writes downloaded data to the ./download directory.

- **Step 3 - Process**: Transforms downloaded data into a brick/smrt_dataset.parquet file.
    - **Dependencies**: ./download - runs if the download changes.
    - **Output**: Stores ./brick/smrt_dataset.parquet.

The SMRT brick also includes a ./devcontainer, enabling developers to easily set up a functional environment for building and publishing the brick. This is beneficial for new developers, tutorials, and runtime environments.

In short, every BioBrick is a git repository with a DVC pipeline written to manage brick creation and updates. Today, the BioBricks.ai system provides a normalized storage and installation system for publishing and depending on bricks. Eventually, the BioBricks.ai system will monitor, update, and store brick data whenever primary sources change.

## 3. Data Formats

Currently, BioBricks.ai supports three primary data types: Parquet, SQLite, and HDT (Header, Dictionary, Triples). These formats were chosen for their specific advantages in handling different types of data and supporting various use cases: We refer to Parquet and sqlite bricks as 'tabular-bricks' and HDT as 'triple-bricks'. The system can distribute any serializable data format, but these formats are preferred; features specially built on these data types may be implemented in the future.

**Parquet:** BioBricks.ai supports Parquet for its compression capabilities, which significantly reduce the size of data files. Parquet also supports partitioning such that one large table can be partitioned into many smaller Parquet files. Compression and partitioning are important for network efficiency, as data can be partitioned into smaller files that are faster to download in parallel. Partitioning is also important for distributed computing systems like Spark and Dask, which are often used to process BioBricks.ai assets (19,20).

**SQLite:** SQLite is used within BioBricks.ai for its robust indexing capabilities and self-contained nature. This simple, serverless database system makes it easy to manage. Its indexing features facilitate quick data retrieval, which is beneficial for operations that require fast access to data (21,22). SQLite's portability and ability to handle complex queries make it an ideal choice for researchers who need to perform detailed data exploration without the overhead of a full database management system.

**HDT (Header, Dictionary, Triples):** BioBricks.ai adopts HDT for managing semantic knowledge graphs. HDT optimizes the storage and querying of RDF (Resource Description Framework) datasets by compressing RDF data and organizing it effectively. This structure supports efficient graph operations and accelerates both data loading and complex querying (23,24). HDT is particularly valuable for projects that involve linked data or require semantic reasoning capabilities.

The Parquet, SQLite, and HDT formats were chosen over others due to their balance of efficiency, flexibility, and widespread support in data science tools and libraries. While BioBricks.ai can distribute any serializable data format, these three formats are preferred for their optimal performance in various data processing scenarios. The system may implement special features built on these data types in the future, further leveraging their unique strengths.

## 4. Results
### 4.1. Capabilities and Performance

Today, BioBricks.ai integrates over ninety biological and chemical datasets. Below are two examples of use cases showing how BioBricks.ai can help with research.

## 4.2. Use Cases
### 4.2.1. Accelerate Data Science

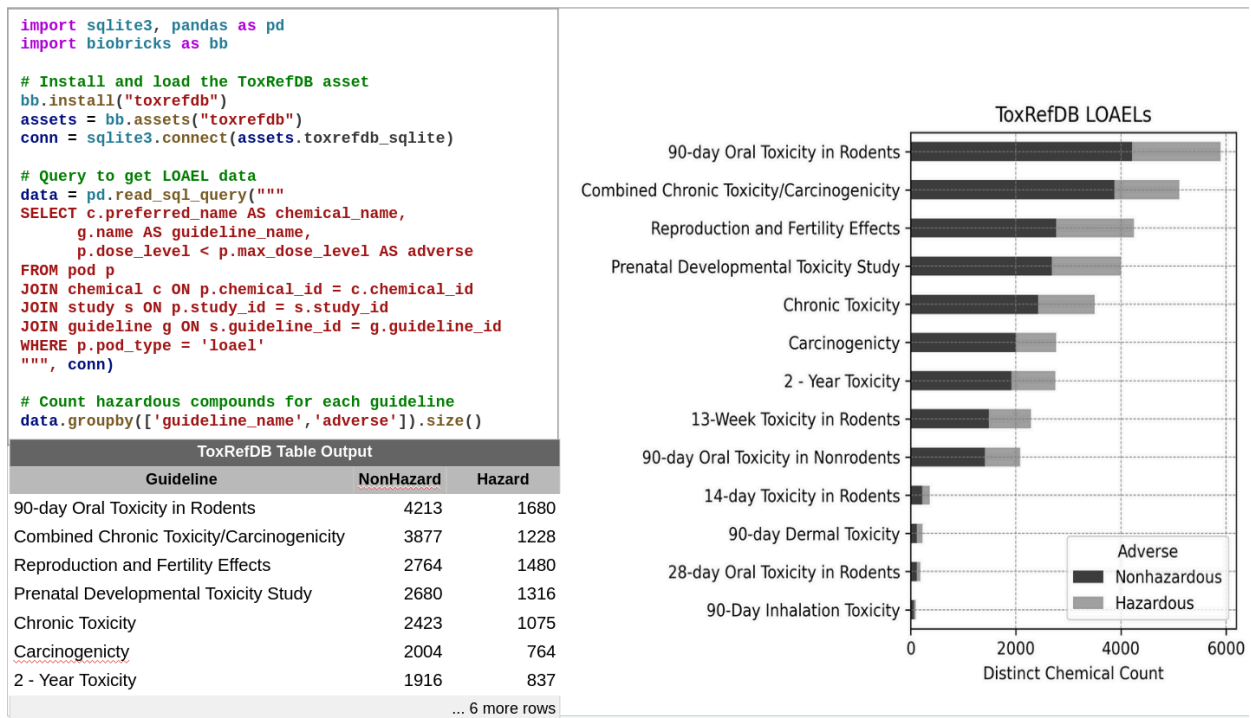

**Figure 1 Top left** - A code example to install, load, and analyze ToxRefDB data. **Bottom left** - the result of running the code example. **Right -** tabular data in bar chart form.

Figure 1 gives an example of a simple analysis workflow. It demonstrates a simplified analysis of the ToxRefDB asset. The ToxRefDB brick is an SQLite asset within BioBricks, containing mammalian toxicity data based on regulatory guidelines useful for chemical risk assessments. The primary source for this data is hosted at Clowder (25). Figure 1 demonstrates how a user can quickly install ToxRefDB, load the provided SQLite asset, and then write a simple query. In this case the query groups compounds into "Hazard" and "NonHazard" based on whether or not the No Observed Adverse Effect Level (NOAEL) is lower than the maximum tested dose.

Biobricks.ai ensures workflow reproducibility and clarity, as it simplifies data loading and traces data versions. Additionally, code written to extract data from the source can now be more easily reused, as it is all contained within an open-source brick - eg. [github.com/biobricks-ai/toxrefdb](github.com/biobricks-ai/toxrefdb). If there is a problem with this brick, an issue can be created and resolved, and an updated version can be published. Most primary sources do not have such a formal structure for communicating about issues and issue resolutions. BioBricks.ai also improves the reproducibility relative to the solution of manually downloading data (via browser clicks, or perhaps sharing institution filesystems),

Using BioBricks makes the resulting code more portable. New developers can simply run the provided code (after installing BioBricks.ai), to reproduce the analysis, all in one step. The python notebook at github.com/biobricks-ai/toxrefdb/blob/main/example.ipynb can be run directly after cloning the repository, or, for tutorial purposes, users can drop into a GitHub codespace (also supported on many BioBricks.ai repos) and run the notebook themselves.

This is a relatively simple example, but it helps to anchor the value supported by BioBricks.ai. BioBricks.ai replaces the time-consuming task of writing ETL pipes and also normalizes the process, so users developing code at different times and places don't perform redundant data-loading work. After this framework is in place, more possibilities are available, providing value through the combination of many data assets.

### 4.2.2. Create Novel Data Assets With Dependencies on Other Assets

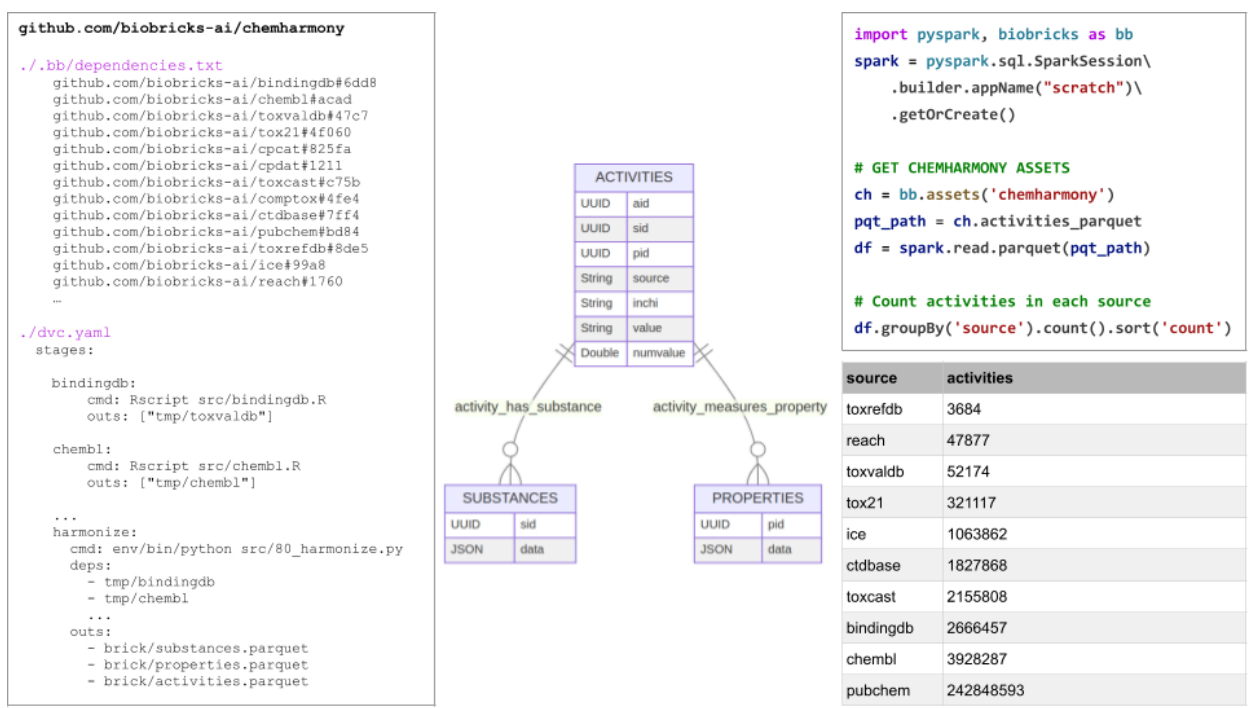

**Figure 2 Left** - truncated versions of the (1) .bb/dependencies.txt and (2) dvc.yaml file in the ChemHarmony BioBrick. **Center**, the 3-table schema of ChemHarmony, a simple chemical activities dataset with a substances, properties, and activities table. **Right** shows how to count activities by source by installing the ChemHarmony brick and using it with Apache Spark with the resulting table in lower right.

BioBricks.ai is useful for creating new assets that depend on one or more other bricks. The ChemHarmony brick, github.com/biobricks-ai/chemharmony, combines many chemical-property data assets into a single, simplified asset primarily created for modeling chemical properties.

The ChemHarmony project is designed to integrate chemical-property values from various databases into a unified system. The database is structured into three main tables: substances, properties, and activities. Each activity links a substance to a property with an assigned value, either binary (indicating positive or negative) or numerical (such as binding affinity or LD50 values), facilitating a quick assessment of chemical characteristics. As shown in Figure 2, the properties table contains the property ID (pid) column and a JSON column containing metadata describing the property. The substances table contains substance ID (sid) and a JSON column describing substance-metadata provided by the substance source. The activities table connects the sid and pid; it also provides structural information such as SMILES and InChi to make it easier to build downstream modeling bricks. The chemharmony code contains scripts to process every source database into the substances, properties, and activities tables thus reducing many complex heterogeneous tabular schemas into one simple schema.

BioBricks.ai supports ChemHarmony by providing the infrastructure and tools necessary to integrate chemical-property data from various sources into a unified database. The databases in ChemHarmony include ChemBL, eChemPortal, ToxValDB, Tox21, CPCat, CPDat, ToxCast, CompTox, CTDbase, PubChem, QSAR Toolbox, BindingDB, ToxRefDB, ICE, and REACH, with several more additions in progress, including RTECS, PubChem-annotations, and Clintox(25–41). ChemHarmony includes 117 million chemicals and 254 million activities, with 4,026 major properties having over 1,000 activities each, including 246 million activities and 4.1 million chemicals in these major properties.

ChemHarmony has already gone through several revisions, with more to come. The BioBricks system made it easy for a team of people working on this asset to collaborate without worrying about synchronizing data dependencies between developer environments. This also means that releases of the ChemHarmony asset have unambiguous dependencies on upstream databases. This same system can be used to indicate when ChemHarmony needs updating and trigger a rebuild. Figure 2 - Left shows a truncated version of the .bb/dependencies.txt ChemHarmony file. This file is built when a user runs biobricks init within a BioBricks.ai repository. It references the git repo and commit hash of each asset that the repository depends on. When a user runs biobricks pull within this repository, all of the bricks they need, but do not currently have, are installed. This allows all ChemHarmony developers to maintain a standardized environment.

When dependencies change, ChemHarmony can be updated easily through manual modification of the .bb/dependencies.txt file or by running biobricks add <brick>, ensuring that all components are up-to-date without disrupting the workflow. All developers work with the same version of the data, thanks to the standardized management of dependencies and data integration provided by BioBricks.ai. This consistency improves collaboration and reduces errors.

### 4.2.3. BioBricks.ai accelerates the exploitation of existing knowledge

Scientific literature provides a valuable source of information about the relationships between biological and chemical entities, which can, for example, support drug repurposing or the discovery of unforeseen drug adverse events. A significant portion of this literature is gathered in PubMed in the form of publicly available abstracts. However, regular changes to the API, as well as a rate-limiting process, make it challenging to analyze the entirety of the corpus. Downloads are possible in the form of FTP bulk transfers, but these then need to be stored in a dedicated platform, which might be disconnected from the analysis environment.

The pubmed biobrick allowed us to apply a simple Apache spark- and spaCy-based NLP pipeline to all of PubMed abstracts, extracting chemical entities, adversities and their potential causal relationships. The relevant methods are discussed (42) and a script is available at [github.com/ontox-hu/pubmed-entox](github.com/ontox-hu/pubmed-entox), this script also provides a simple demonstration of how to use the pubmed brick. The pubmed brick itself is 66.3 gigabytes, it distributes a single asset, pubmed_parquet, which is a large Parquet table containing PubMed IDs and MEDLINE metadata in a JSON-formatted string column. This BioBrick is updated in an append-only manner, allowing users to download only the new data without needing to re-download previously obtained parts of the table; this is handled automatically by the biobricks install mechanism. Writing BioBricks to handle sources with frequent small updates can be challenging, but in the future, BioBricks may leverage other "git-for-data" approaches, such as Delta Lake, which can more efficiently handle updates to large Parquet files.

5. **Discussion**
5.1. **Comparison with Existing Technologies**

The current landscape of public data asset management in the life sciences involves many different data sources, each using their own ad-hoc distribution system, including APIs, FTP, custom data formats, and custom packages. Data users face challenges due to the need to write custom code for each source, often redundantly with others. For instance, a researcher studying drug interactions might need to navigate NCBI's interface for genomic data, use ChEMBL's API for chemical structures, and parse custom formats from toxicology databases. Two common approaches to addressing these problems are federated data approaches and harmonization assets.

**Federated data approaches** such as the Semantic Web require each data source to adopt standardized ontologies, which are comprehensive frameworks for data representation. This standardization facilitates normalized querying across different data sources, allowing users to access and integrate diverse datasets through a common interface. However, the high coordination cost and complexity involved in achieving this standardization can be significant barriers for data developers. While high effort, ultimately, federation can be a successful approach, and great strides have been made towards unifying many data sources. For instance, the PubChem RDF (43) successfully harmonizes data from MeSH, UniProt, PDB, ChEBI, Reactome, ChEMBL, and Wikidata and provides a SPARQL endpoint that can reliably query data across these graphs.

**Harmonization assets** involve groups building new assets to simplify and reduce heterogeneity in the life-science data ecosystem. Examples include PharmacoDB, Harmonizome, ROBOKOP, Hetionet & SPOKE, ComptoxAI, Bioteque, Drugbank, and DRKG. These sources take upstream data sources and develop code to extract, transform, and load those sources into new data assets. While this approach reduces coordination costs by moving the burden of data transformation to the developer of the new harmonized asset, it also adds complexity to the data ecosystem. Each new asset creates maintenance costs, and the proliferation of such assets often results in redundant efforts.

Table 2 - Comparison of features of BioBricks.ai with other approaches to data distribution in life sciences

| Feature | BioBricks.ai | Federated Data Approaches (semantic web) | Harmonization Assets | Traditional Data Repositories |
|---|---|---|---|---|
| Data Integration | Supports multiple data types and sources | When standards are adopted | Within specific domains | Often siloed |
| Standardization | Standardized access, flexible data formats | Requires coordination on ontologies | Within asset scope | Varies widely |
| Ease of Use | Package manager-like system | Requires specialized knowledge | Depends on asset | Often requires manual navigation |
| Reproducibility | Version control built-in | Depends on implementation | Within asset scope | Often lacks versioning |
| Scalability | Designed for large datasets | Depends on infrastructure | Asset-dependent | Often limited by original design |
| Community Contribution | Open-source model | Depends on governance | Often centrally managed | Usually closed systems |
| Data Update Frequency | Can be real-time | Depends on participants | Often periodic releases | Varies widely |
| Interoperability | Common access method for diverse data | When standards are adopted | Within asset scope | Often requires custom integration |
| Learning Curve | New system, but designed for ease of use | Requires understanding of complex standards | Asset-specific knowledge needed | Varies - Often high for each new source |
| Cost Efficiency | Reduces redundant work | High initial investment | Reduces some redundancy | Often leads to redundant work |
| Flexibility for New Data Sources | Can easily add new 'bricks' | Requires adherence to existing standards | Often limited to predefined scope | Can add, but often in isolation |
| Support for AI/ML Applications | Designed with AI/ML needs in mind | Depends on data quality and format | Often designed for specific AI/ML tasks | Often requires significant preprocessing |

Table 2 breaks comparisons down into a set of features compared across BioBricks, semantic-web based federated data approaches, harmonization assets, and the current ecosystem of independent databases. The ratings of the performance of each approach in each key feature (Green = strongly supported, Yellow = moderately supported or problematic, Red = not well supported or highly problematic) are generalizations and may vary depending on specific implementations or use cases.

Neither of the aforementioned approaches addresses the situation where a data user wants to use a new, unsupported data source. The data user must once again write code to ETL the new source in a way that is redundant with other users of the source.

BioBricks.ai proposes a solution by providing a collection of git repositories with ETL code. This approach addresses several key issues. BioBricks.ai reduces coordination costs by allowing data publishers to contribute data in isolation from other data publishers with relatively loose constraints. BioBricks.ai also enables reusable ETL pipelines, significantly reducing the redundant work of writing ETL code for new data sources.

A useful analogy is to describe BioBricks.ai as a package manager for data. By providing a standardized method for namespacing and accessing data assets, BioBricks eases data reuse. This approach is analogous to how package managers like npm for JavaScript or pip for Python have streamlined software development by providing a central repository for code dependencies. Furthermore, BioBricks.ai leverages open-source collaboration principles, using platforms like GitHub to enable community-driven improvements and rapid issue resolution.

Rather than being an alternative to existing technologies, BioBricks can serve as a data layer that accelerates the creation of semantic web applications and decreases the redundancy in the development of harmonization assets. BioBricks.ai demonstrates its functionality as a data layer for the semantic web as part of the NSF Prototype Open Knowledge Network (44). The presented ChemHarmony use case demonstrates that BioBricks.ai brick dependencies can aid in the construction of harmonization assets.

## 5.2.   Limitations and Future Work

While BioBricks.ai offers significant advantages for public health data management, it faces several practical limitations. Here, we discuss these limitations and propose potential solutions for future development:

**Large Datasets**: Downloading multi-terabyte–sized assets is impractical for many users. **Future Work:** We plan to allow users to query and interact with large datasets without requiring full downloads. This could involve implementing a cloud-based query engine, developing a data chunking system, or creating a caching system for frequently accessed data.

**Pricing:** While BioBricks.ai does help cache data, frequent data transfers are still costly.
**Future Work:** To further optimize costs, we will implement a tiered access system, develop partnerships with cloud providers, and allow distributed hosting of bricks.

**Complexity:** New users can sometimes struggle to set up the system.
**Future Work**: To simplify the user experience, we plan to deploy remote developer environments that allow new users to quickly get started with bricks without needing to set up and configure a local system.

**Data Quality:** Ensuring consistent data quality across diverse sources can be challenging.
**Future Work**: We propose implementing automated data quality checks, developing a standardized metadata schema, and creating a community-driven review system for bricks.

**Real-time Updates:** Keeping all bricks up-to-date with their primary sources can be challenging.
**Future Work**: We plan to develop an automated system to monitor primary sources for changes, implement a versioning system, and create notification systems for critical updates.

**Tracking:** Some users might be interested in downloading information about particular bricks.
**Future Work:** Add logging to the packages so we can track which bricks are being used.

**Integration with Existing Workflows:** Users may face challenges integrating BioBricks.ai into their data analysis pipelines.
**Future Work**: To facilitate easier integration, we propose developing APIs and SDKs for popular programming languages, creating adapters for common bioinformatics tools, and collaborating with other data standardization efforts.

**Scalability:** System performance may struggle as the number of bricks and users grow.
**Future Work**: We plan to implement a distributed architecture, optimize data storage and retrieval mechanisms, and develop load balancing systems.

By addressing these limitations in our future work, we aim to make BioBricks.ai an even more powerful and user-friendly platform for public health data management. We welcome community collaboration in tackling these challenges and shaping the future direction of BioBricks.ai.

Two future directions under active development include developer environments (now supported on many bricks) and LinkML descriptions.

**Developer Environments:** To address limitations, BioBricks.ai intends to build developer environments with built-in access to the full set of available BioBricks.ai assets. This approach largely mitigates the limitations of large data assets and hides the complexity of installing and managing a brick library. Developers can start using data immediately in a fully functional developer environment, thus facilitating a smoother onboarding process.

**LinkML:** LinkML is a versatile modeling language that allows the creation of schemas in YAML (human-readable data serialization language) to describe data structure. BioBricks.ai will use LinkML to create meta-models for data validation and description. LinkML provides a metadata description of the data distributed by a brick. It will describe each table and column for tabular data (or Parquet files). Each BioBrick will have a LinkML schema specifying the structure and content of its files. It makes curating, mapping, ingesting, and organizing data much easier. It supports a range of applications from simple tasks such as describing spreadsheet columns to the creation of complex interlinked schemas. LinkML is compatible with various frameworks. LinkML offers many features that could enhance BioBricks.ai, like generating schema, validation, and linting of schemas, data conversion, validation, and programmatic manipulation of schemas. (LinkML, 2021-2024). This approach will improve data clarity and consistency, using Linked Data principles for better data management.

## 6. Conclusion

BioBricks.ai represents a significant advancement in the management and distribution of biomedical research and public health data, offering a solution to the longstanding challenges of data fragmentation, accessibility, and reproducibility in the life sciences. By extending the principles of part reuse and standardization to public health data management, BioBricks.ai is poised to accelerate scientific discovery and innovation across various fields, including drug discovery, toxicology, biochemistry, and public health research.

## 7. Additional Information

**CLI Package:** BioBricks.ai command line interface is on PyPI pipx install biobricks

**Client Packages:** The R and python packages can be installed from cran and PyPI

**Source Code:** The BioBricks.ai command line interface is [github.com/biobricks-ai/biobricks](github.com/biobricks-ai/biobricks)

**Operating System:** Biobricks.ai supports Windows, Mac, and Linux.

**Usage Restrictions:** BioBricks.ai is open source with an MIT license.

Please read more detailed used cases on [https://insilica.co/posts/](https://insilica.co/posts/).

We used ChatGPT in the writing process.

## 8. Funding

BioBricks.ai is a product of Insilica LLC and has been funded by:

**NSF SBIR 2012214**
Advanced Cancer Analytics Platform For Highly Accurate and Scalable Survival Models to Personalize Oncology Strategies.

**NIH SBIR 1R43ES033851-01**
ToxIndex-CPG: Machine learning driven platform integrating a hazard susceptibility database to quantify chemical toxicity factors, predict risk levels and classify biological responses.

**EU Horizon 2020 963845**
Ontology-driven and artificial intelligence-based repeated dose toxicity testing of chemicals for next generation risk assessment.

**NIH SBIR 1R43ES036069-01**
BioBricks-Env: AI driven, open-source, modular, composable platform to organize, store, retrieve, extract, and integrate environmental health & risk related data.

**NSF Award #2333728**
BioBricks-OKG An Open Knowledge Graph For Cheminformatics And Chemical Safety.

## 9. Acknowledgements

We gratefully acknowledge the support of our collaborators and the developers of the BioBricks.ai platform. Special thanks to the developers of individual bricks, the Ontox project community for their valuable feedback throughout the project lifecycle, and our funding and communication support from NIEHS, NICEATM, and NSF.

## 10. Conflict of Interest

The authors declare the following potential conflicts of interest regarding the research and publication of this paper: BioBricks is a product developed by Insilica LLC, and many of the authors are employees of Insilica LLC. As such, there may be a perceived or real financial interest in the outcomes of the research and the development of BioBricks. The authors affirm that their contributions to the research and the manuscript were conducted with scientific integrity and without bias influenced by their association with Insilica LLC.

## 11. Author Contributions

**Yifan Gao**
Gao has written several Bricks and completed the analyses and documentation of the product in this paper.

**Marc A.T. Teunis**
A member of Ontox, Teunis is a frequent user of the BioBricks platform and has provided substantial guidance throughout the development of the platform.

**Marie Corradi**
A member of Ontox, Corradi is a frequent user of the BioBricks platform and has provided substantial guidance throughout the development of the platform.

**Suliman Sharif**
Sharif is the newest biobricks.ai developer. He has developed ~5 bricks at this point. He is an Insilica employee.

**Ajay Chatrath**
Chatrath was one of our first brick developers and informed some of the first decisions to create the platform.

**Karamarie Fecho**
Fecho provides guidance in the BioBricks-OKG project, an open knowledge graph for chemical safety. She has done application testing and identified issues for product improvement.

**John Shaffer**
Shaffer was one of the first BioBricks developers. He also built status.biobricks.ai and informed the architecture of the platform. He is a developer at Insilica LLC.

**Alexandre Borrel**
Borrel has built a few bricks at the early project stage and has been one early tester of BioBricks.ai.

**Ben Lieberman**
Lieberman has built many of our bricks and is a developer at Insilica LLC.

**Zakariyya Mughal**
Mughal is current technical lead of BioBricks.ai.

**Jose Jaramillo**
Jaramillo wrote the first implementation of BioBricks. He was the original architect of the pypi biobricks package, helped to formulate the original ideas and is an active member of the community.

**Alexandra Maertens**

Maertens is a member of Ontox and provided editorial suggestions to the paper.

**Thomas Hartung**
Hartung has provided guidance throughout the lifetime of the BioBricks.ai project. He has been an invaluable stakeholder for the project through efforts around communicating the ideas and use cases for the platform.

**Nicole Kleinstreuer**
Nicole Kleinstreuer has provided guidance and support around understanding the place for BioBricks.ai in the wider informatics ecosystem and has provided feedback for selecting which bricks to develop, how to update existing bricks, and communicated the platform to other members of the community.

**Thomas Luechtefeld**
Luechtefeld formulated the platform, drove the construction and architectural design, contributed to many bricks.

All authors contributed to the manuscript and approved submission of the final draft.